\documentclass[aps,pre,twocolumn,showpacs,superscriptaddress]{revtex4-1}

\usepackage{amsmath}
\usepackage{amssymb}
\usepackage{graphicx}
\usepackage{lineno}
\usepackage{amsmath}

\begin{document}

\title{Signal Manifestation Trade-offs in Incoherent Feed-Forward Loops}

\author{Tarunendu Mapder}
\email{mtarunendu@yahoo.com}
\affiliation{Department of Chemistry, Indian Institute of Engineering Science and Technology, Shibpur, Howrah 711103, India}



\begin{abstract}

Signal processing in biological systems is delicately executed by specialised networks, which are modular assemblies of network motifs. The motifs are independently functional circuits found in enormous numbers in any living cell. A very common network motif is the feed-forward loop (FFL), which regulates a downstream node by an upstream one in a direct and an indirect way within the network. If the direct and indirect regulations go antagonistic, the motif is known as an incoherent FFL (ICFFL). The current study is aimed at exploring the reason for the variation in the evolutionary selection of the four types of ICFFLs. As comparative measures, I compute sensitivity amplification, adaptation precision and efficiency from the temporal dynamics and mutual information between the input-output nodes of the motifs at steady state. The ICFFL II performs very efficiently in adaptation but poor in information processing. On the other hand, ICFFL I and III are better in information transmission compared to adaptation efficiency. Which is the fittest among them under the pressure of natural selection? To sort out this puzzle, I take help from the multi-objective Pareto efficiency. The results, found in the Pareto task space, are in good agreement with the reported abundance level of all the types in eukaryotes as well as prokaryotes.

\end{abstract}

\maketitle

\section{Introduction}

The essence of biological signal transduction is to sense external signal and proper manifestation of it. The sensitivity of a bio-signaling system is defined as the magnified response to every minute change in the exterior \cite{Shoval2010, Macnab1972}. But it is not enough to ascribe a signalling machinery. Apart from the precise response to an external stimulus, the regulation of the system becomes obligatory to motivate its own features in accord with sustaining stimulations. The motivation to restore itself after sensing the environmental variation is known as adaptation \cite{Koshland1982}. A signalling pathway is said to be good adaptive when its response to a persistent stimulation is transient, and a quick restoration ensues the pre-stimulus basal level. The goodness of adaptation depends upon the return of the response to the near-basal level after a period of the response. From the level of microscopic cell signalling to nerve conduction, a precise adaptation is unavoidable to keep the sensing outfit functional for further proceedings. Adaptation is observed in bacterial chemotactic system \cite{Yi2000, Hansen2008, Endres2006}, osmotic response system \cite{Miller1986}. Higher eukaryotes like mammals also have adaptive features in olfactory receptors \cite{Reisert2001}, and hormone receptors \cite{Stadel1983}. The definite adaptation in a biological network may be obtained by recurrent or random metabolic regulation, which may be favoured in the selection pressure of convergent evolution. When a system like the chemotaxis in {\it E. coli} fails to adapt precisely, the robustness of the system also decreases \cite{Hansen2008, Endres2006}. On the other hand, at the sub-cellular level, a dynamical network is probabilistic due to the rare molecular population. The generated fluctuations make the network noisy \cite{Tsimring2014}. Hence, to decipher the association between the network input and output, one has to decrypt the fidelity through the quantification of mutual information \cite{Tostevin2010}.

In the current study, I aim to focus on few adaptive biochemical network motifs, some of which enjoy favour over the others in the continuous evolutionary pressure \cite{Alon2006}. I choose four classes of incoherent feed-forward loops (ICFFL) ( see Fig.~\ref{fig1}), because they have tendency to produce adaptive responses and they also have uneven abundance in biological regulatory networks \cite{Mangan2003a}. There are several theoretical and experimental reports on FFL motifs playing crucial role in gene regulation and signal transduction \cite{Mangan2003a,Mangan2003b,Mangan2006,Kalir2005,Kaplan2008,Dekel2005,Milo2002,Kremling2008,Wall2005,Guo2009,Ghosh2005,Bose2004}. The FFL motifs have a role in fold change detection, which is essential for information flow in noisy networks \cite{Goentoro2009, Shoval2010}. The coherent FFLs sometimes transduce only persistent stimuli, not the transients. The CFFLs are always activating in nature. As a result, they perform good in noise reduction and type-1 CFFL is least noisy \cite{Ghosh2005}. But in case of ICFFL noise does not get reduced and no conclusion can be made in this way \cite{Ghosh2005}. Hence, the selection procedure for the ICFFLs is something different. It is true that noise in gene regulation networks has relevant effect on the evolutionary selection. But it is not the only decisive aspect. The sensitivity amplification and precise adaptation also are as important as noise utilisation.

The present analysis of the ICFFL motifs is based on their performance in sensitivity, information processing and adaptation efficiency in a range of graded induction level. The sensitivity and adaptation are quantified from the mean-field time trajectory at every step increment of the induction and the information is measured at steady state. To remark on the trade-off among the aforesaid tasks in the ICFFLs, I use Pareto optimality. The adaptation efficiency and mutual information are projected in the task performance space to check the optimality, because there are few such universal tasks to complete with a plenty of molecular organisations on the basis of their priority in cellular organisms.

\section{The Model}
\subsection{Incoherent feed-forward loops}

\begin{figure*}[h]
\begin{center}
\includegraphics[width=0.6\linewidth,angle=0]{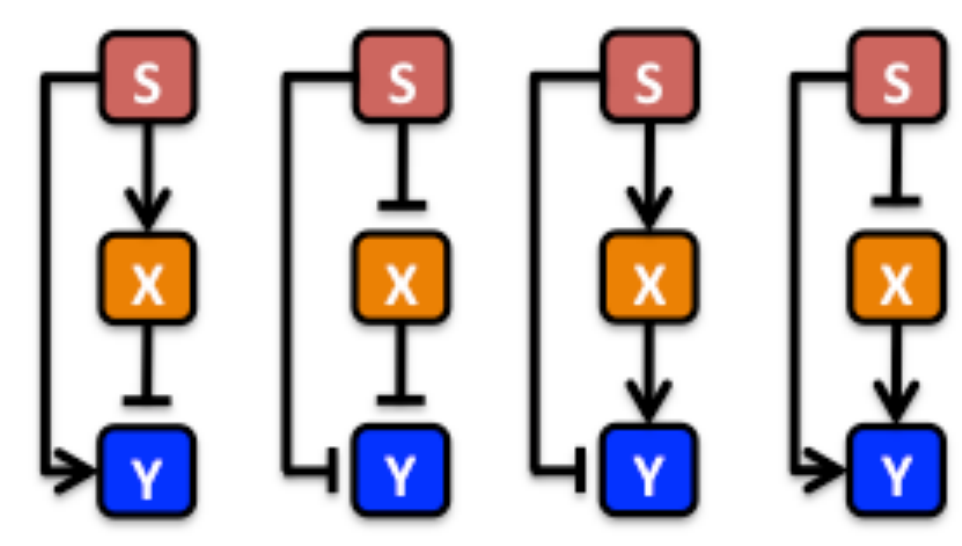}
\caption{\textbf{The four types of incoherent feed-forward motifs.}
From the left: type I, type II, type III and type IV respectively. Here $S$, $X$ and $Y$ in all the motifs are the network components or nodes, connected with the edges. The arrow-headed edges shows activation and blunt-headed edges stand for repression. $S$ and $Y$ are respectively the input and output of the motifs.}
\label{fig1}
\end{center}
\end{figure*}

The regime of the present study starts with the adaptive and strategic characterization of few gene transcription regulatory network (GTRN) motifs. All the motifs are exhibited in Fig.~\ref{fig1}. Here, I focus on the four different classes of incoherent feed-forward loops (ICFFL) \cite{Alon2006}. All the GTRN motifs considered in Fig.~\ref{fig1} have a network input S and an output Y. A midway agent X indirectly conveys the instruction coming from S; otherwise, Y is directly monitored by S. The common feature of the four types is the regulation of Y through S and X are in opposite phase, i.e., incoherent. From Fig.~\ref{fig1} and Table II one can see that S has an excitatory role on X in type I and III and is inhibitory in the other two categories (type II and IV). The direct S-Y regulation is positive in type I and IV, and negative in type II and III where the indirect X-Y control is positive in type III and IV in contrast to the case of I and II.  Interestingly in type II motif, all the regulations are inhibitory. I will discuss in results that it may help in favour of perfect adaptation. All the network motifs considered in Fig.~\ref{fig1} can be mathematically presented in set of ordinary differential equations as:
\begin{eqnarray}
\label{eq1}
\frac{ds}{dt} &=& k_{0} + k_{1}I - \tau^{-1}_s s, \\
\label{eq2}
\frac{dx}{dt} &=& k_{2}f_x(s) - \tau^{-1}_x x, \\
\label{eq3}
\frac{dy}{dt} &=& k_{3}f_{y}(s)g_{y}(x) - \tau^{-1}_y y.
\end{eqnarray}
\noindent
where, $k_{0}$ and $k_{1}$ are the basal level and stimulated synthesis rate parameter of S and $k_2$, $k_3$ are the maximum synthesis rate constants for X and Y, respectively. $\tau_s$, $\tau_x$, $\tau_y$ are the lifetimes for S, X and Y, respectively. So I can consider $\tau^{-1}_s$, $\tau^{-1}_x$ and $\tau^{-1}_y$ as the degradation rate constants of them accordingly. The details of the kinetic scheme and {\textit f($\cdot \cdot$)}s are tabulated in Table I and Table II.

\subsection{Sensitivity and adaptation}

\begin{figure*}[h]
\begin{center}
\includegraphics[width=0.6\linewidth,angle=0]{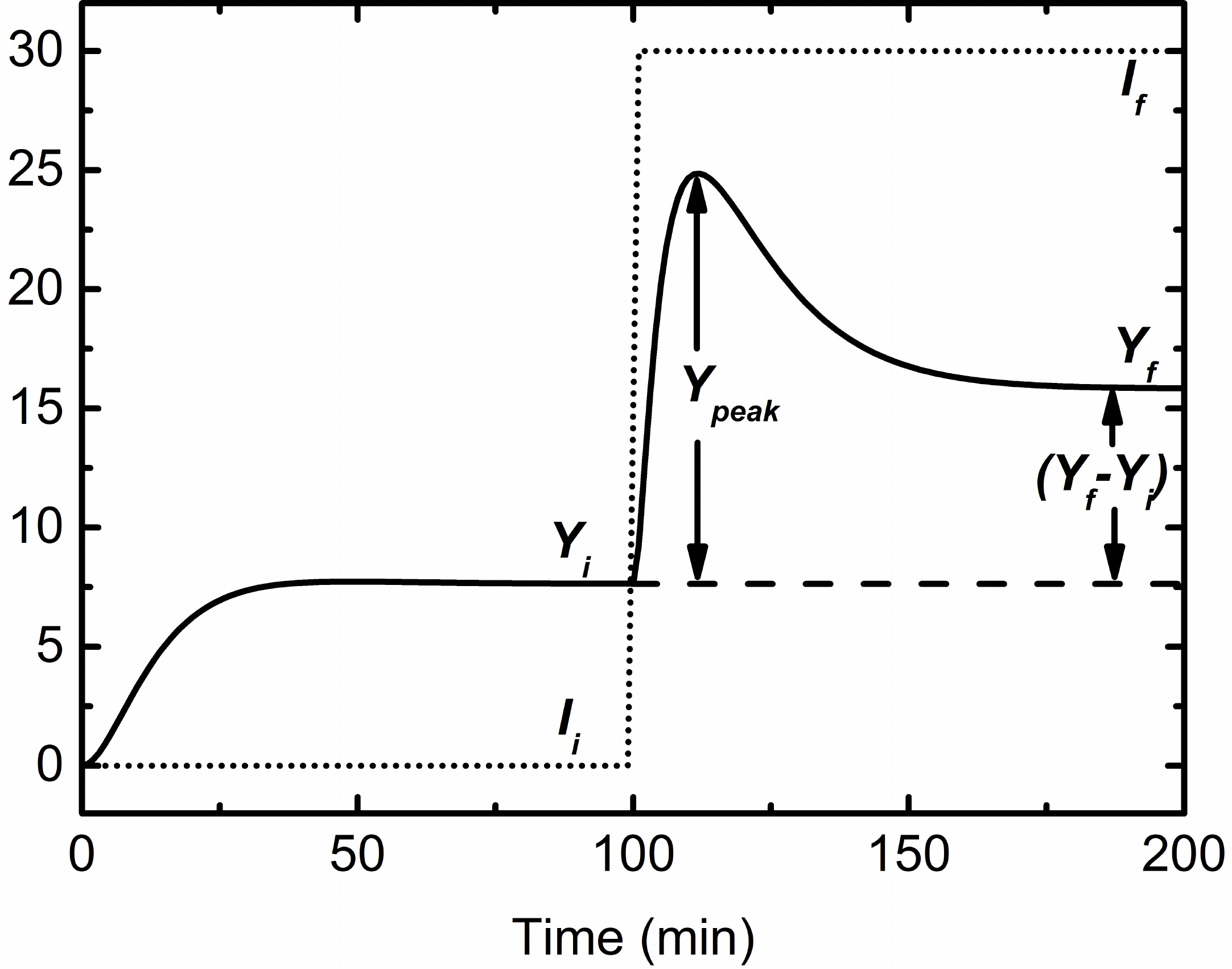}
\caption{\textbf{Schematic diagram for sensitivity and adaptation.}
 For near zero inducer level, $I_i$, the level of the output, $Y_i$ saturates at moderately low value, which can be called as pre-stimulus state. On triggered induction, $I_f$, the output reaches a sharp oversoot, $Y_{peak}$ and reaches a lower steady state, $Y_f$. The conditions of better sensitivity, and perfect adaptation are $Y_i << Y_{peak}$ and $Y_f-Y_i \approx 0$ respectively. On the fulfillment of both the conditions, the adaptation efficiency will increase. (The numbers in both the axes are representative).}
\label{fig2}
\end{center}
\end{figure*}

Here, I am going to discuss the sensitivity amplification, the adaptation precision and efficiency for the four ICFFLs in a comparative study. As the living world survives through biochemical communications in a particular niche, the communication machinery has to be smart enough to respond to a particular stimulus precisely. The nature of the response to a sustaining stimulus may be sharp sensitive, efficiently or perfectly adaptive \cite{Briat2016}. In the present study, the signal from the environment is $I$. $I_i$ and $I_f$ in Fig.~\ref{fig2} stand for the basal and induced level of the stimulus respectively. The increment of $I$ has been considered as a step elevation. The model output, Y also shows basal and induced steady state expression as $Y_i$ and $Y_f$. For the current adaptive model, Y follows a quick overshoot $Y_{peak}$ before reaching the delayed steady state. The sensitivity amplification ($A_S$) can be written symbolically as \cite{Goldbeter1984, Huang1996}:
\begin{eqnarray}
\label{eq4}
A_S &=& \bigg| \frac{(Y_{peak}-Y_i)/Y_i}{(I_f-I_i)/I_i} \bigg|,
\end{eqnarray}
\noindent
which explains the maximum variation in the output response with respect to the relative variation in input stimulus. The mathematical definition of adaptation precision ($\delta$) is \cite{Ma2009, Jia2016}
\begin{eqnarray}
\label{eq5}
\delta &=& \bigg| \frac{(Y_f-Y_i)/Y_i}{(I_f-I_i)/I_i} \bigg|^{-1},
\end{eqnarray}
\noindent
which says about the capacity of a system to minimise the steady state variation in the output response. A biological network is supposed to be efficient if both the $A_S$ and $\delta$ are quite large. A high sensitivity makes a system better detector of every little variation in the environment. On the other hand, a precise adaptation makes the system to recover quickly to its pre-stimulus level. By the assembly of the above two parameters, one can calculate adaptation efficiency as \cite{Jia2016}:
\begin{equation}
\label{eq6}
\eta = 1 - \frac{1}{A_S \delta} = \bigg| \frac{Y_{peak}-Y_f}{Y_{peak}-Y_i} \bigg|.
\end{equation}
\noindent
For an efficient biological system $\eta$ should tend to 1 while it is always less than 1. The utility of this efficiency measure is that it does not depend on the stimulus level directly. 

\subsection{Information flow}

The Shannon's formalism of mutual information is a quantitative measure of the association between two or more elements in a noisy signalling network \cite{Shannon1948}. The current motifs of our interest are performing at constant signal levels, and the information calculations are done at steady state where the fluctuations are Gaussian-type. To quantitate the information flow I need to describe the time-dependent dynamics of the network components in the form of chemical Langevin equations as \cite{Kampen2007, Paulsson2005}:
\begin{eqnarray}
\label{eq7}
\frac{ds}{dt} &=& k_{0} + k_{1}I - \tau^{-1}_s s + \xi_s,\\
\label{eq8}
\frac{dx}{dt} &=& k_{2}f_x(s) - \tau^{-1}_x x + \xi_x,\\
\label{eq9}
\frac{dy}{dt} &=& k_{3}f_{y}(s)g_{y}(x) - \tau^{-1}_y y +\xi_y.
\end{eqnarray}
\noindent
The fluctuations associated with the different network components are $\xi_s$, $\xi_x$ and $\xi_y$ respectively. To solve the set of coupled Langevin equations, I opt for the linear noise approximation, which performs satisfactorily in weakly non-linear kinetic systems \cite{Maity2014,Elf2003,Swain2004,Paulsson2004}. At this point, a generalised matrix form makes the calculations tractable. I define matrices, ${\textbf Z}=(s,x,y)$ and $\boldsymbol{\Xi}(t)=(\xi_s,\xi_x,\xi_y)$ and rewrite the linearized Langevin equation at steady state as \cite{Kampen2007,Gardiner2009}:
\begin{equation}
\label{eq10}
\frac{d \delta {\textbf Z}(t)}{dt} = {\textbf J}_{{\textbf Z}= \langle {\textbf Z} \rangle} \delta {\textbf Z}(t) +\boldsymbol{\Xi} (t),
\end{equation}
\noindent
where, $\delta {\textbf Z} (t) = {\textbf Z} (t)- \langle {\textbf Z} \rangle$ is the fluctuations around the steady state, $\langle {\textbf Z} \rangle$. ${\textbf J}$ is the Jacobian matrix calculated at steady state as
\begin{equation}
\label{eq11}
{\textbf J}= 
\left (
\begin{array}{ccc}
-\tau^{-1}_s & 0 & 0 \\
k_2\frac{\partial f_x(s)}{\partial s} & -\tau^{-1}_x & 0 \\
k_3g_y(x)\frac{\partial f_y(s)}{\partial s} & k_3 f_y(s)\frac{\partial g_y(x)}{\partial x} & -\tau^{-1}_y
\end{array}
\right ).
\end{equation}
\noindent
To calculate the variance and covariances from the linearized Langevin equation, I have constructed the Lyapunov matrix equation
\begin{equation}
\label{eq12}
{\textbf J} \boldsymbol{\sigma} + \boldsymbol{\sigma} {\textbf J}^T + {\textbf D} = 0,
\end{equation}
\noindent
where, $\boldsymbol{\sigma}$ is the covariance matrix and ${\textbf D}= \langle \boldsymbol{\Xi} \boldsymbol{\Xi}^T \rangle$ is the diffusion matrix which can be written as
\begin{equation}
\label{eq13}
{\textbf D}= 
\left (
\begin{array}{ccc}
2\tau^{-1}_s \langle s \rangle & 0 & 0 \\
0 & 2\tau^{-1}_x \langle x \rangle & 0 \\
0 & 0 & 2\tau^{-1}_y \langle y \rangle
\end{array}
\right ).
\end{equation}
\noindent
The covariance matrix helps me to calculate the mutual information ${\cal I}(s,y)$ between the input, S and output, Y. Under the consideration of the white Gaussian noise, I may write the form of Shannon's mutual information as \cite{Cover2012}:
\begin{equation}
\label{eq14}
{\cal I}(s,y)= \frac{1}{2}log_2 \bigg( 1+ \frac{\sigma^4_{sy}}{\sigma^2_s \sigma^2_y-\sigma^4_{sy}} \bigg).
\end{equation}
\noindent
Here, $\sigma^2_s$ and $\sigma^2_y$ are the variance of S and Y respectively, and $\sigma^2_{sy}$ is the covariance between S and Y. The quantity $\sigma^4_{sy}/(\sigma^2_s \sigma^2_y-\sigma^4_{sy})$ is known as the measure of fidelity or the signal-to-noise ratio.

To validate the theoretical treatments, a numerical stochastic simulation has been performed. The Monte Carlo simulation based Gillespie algorithm \cite{Gillespie1976, Gillespie1977} can reproduce the analysis of Langevin formalism satisfactorily for the present system. In the numerical evaluation, the variances and covariance have been computed, and mutual information has been quantified using Eq.~\ref{eq14}.

\subsection{Pareto optimality}

For a multitasking biochemical motif, there should be a trade-off among all the different tasks. A single motif is not a specialist in all the tasks but may be in any of them. On the contrary, a motif may exhibit the role of the generalist, which maintains the trade-off between the tasks  \cite{Noor2012}. I want to observe that among the four ICFFL motifs, which one is effectively more generalist than the others in all tasks. The tasks, I am going to discuss here, are the adaptation efficiency ($\eta$) and the mutual information (${\cal I}(s,y)$). A particular network motif will be considered efficient if it is sensitive as well as efficiently adaptive and also informative. In the present study, I am going to assess the fitness of the motifs in the evolutionary trade-off. To test the optimality in the fitness of the four ICFFLs, I have adopted a flavour of the Pareto efficiency \cite{Noor2012,Shoval2012,Sheftel2013}.

The idea of Pareto optimality may achieve popularity when one is in big trouble with a situation of optimising a model satisfying multiple criteria \cite{Pareto1971}. In a multidimensional task space, a point is said to be {\it Pareto efficient}, if a movement from another point towards it can make an improvement in any of the tasks without compromising the quality of other tasks. In the case of biological phenotypes with multiple tasks, the task space overlaps with the trait space. The maxima of each task at the trait space is called the archetype \cite{Shoval2012}. The geometry connecting the archetypes are the Pareto fronts or the Pareto optimal surfaces. The Pareto front is a straight line for space with two tasks, triangle for three tasks and tetrahedron for four tasks and so on. A point within or nearest to the Pareto front is more Pareto efficient than a distal point.

\section{Results and Discussion}

To illuminate the insight on the imparity of selection for the four ICFFLs, I calculate few measurable entities deterministically as well as stochastically. The sensitivity ($A_S$), adaptation precision ($\delta$) and adaptation efficiency ($\eta$) are computed through deterministic approach. Lastly, I measure the mutual information (${\cal I}(s,y)$) between the network input (S) and output (Y) from the stochastic consideration. For simplicity and quantitative comparison, I have assumed all the reaction rate constants to be same for all types of motifs correspondingly (see Table I). Fig.~\ref{fig3} presents (a) $A_S$, (b) $\delta$, (c) $\eta$ and (d) ${\cal I}(s,y)$ sequentially. The lines in four different shades represent four types of incoherent feed-forward motifs (solid for type I, dash for type II, dot for type III and dash dot dot for type IV). To examine the relative fitness of all the motifs, I plot an objective space that suggests the optimality of the ICFFLs according to the available dataset of their abundance \cite{Mangan2003a}.

\subsection{Role of negative regulation on sensitivity amplification}

In the purview of the motif structure and presence of negative regulations, I want to grade the four motifs of the current study. The type I, III and IV have only a single negative control, each at different positions of the circuits. In type I, it is at $X\rightarrow Y$. In type III, the $S\rightarrow Y$ is negative, and $S\rightarrow X$ is negative in type IV. Only the type II ICFFL possess all the three regulations as negative ($S\rightarrow X$, $X\rightarrow Y$ and $S\rightarrow Y$) and it behaves uniquely among the four (The dashed lines in Fig.~\ref{fig3}). It is explained clearly in the trend of sensitivity ($A_S$) in Fig.~\ref{fig3}(a). At low induction level, the ICFFL II is better sensitive than the others and it holds the same quality all over the range of signal strength. On the other hand, at high-level of stimulus, the ICFFL I and IV can produce good sensitivity, but type III loses sensitivity. I want to segment the analysis in two stages of induction, low and high separately, keeping the unique type II apart. The negative controls in ICFFL I, III and IV provide quite similar features at the low signal range, but they get explored distinctly at the high signal.

To observe type I and III analytically, the only difference is in the regulatory functions $f_y(s)$ and $g_y(x)$ of Y. At the same node of these two three-node motifs, only the exchange of negative edge from indirect ($g_y(x)$) to direct ($f_y(s)$) makes the sensitivity dissipate at sufficiently high induction. If I make the negative edge hop at three different combinations among type I, III and IV, a notable comment can be made that the direct $S\rightarrow Y$ regulation dominates over the indirect negative $X\rightarrow Y$ and $S\rightarrow X$ regulations of Y.

\subsection{Quality of adaptation and network topology}

\begin{figure*}[h]
\begin{center}
\includegraphics[width=1.0\linewidth,angle=0]{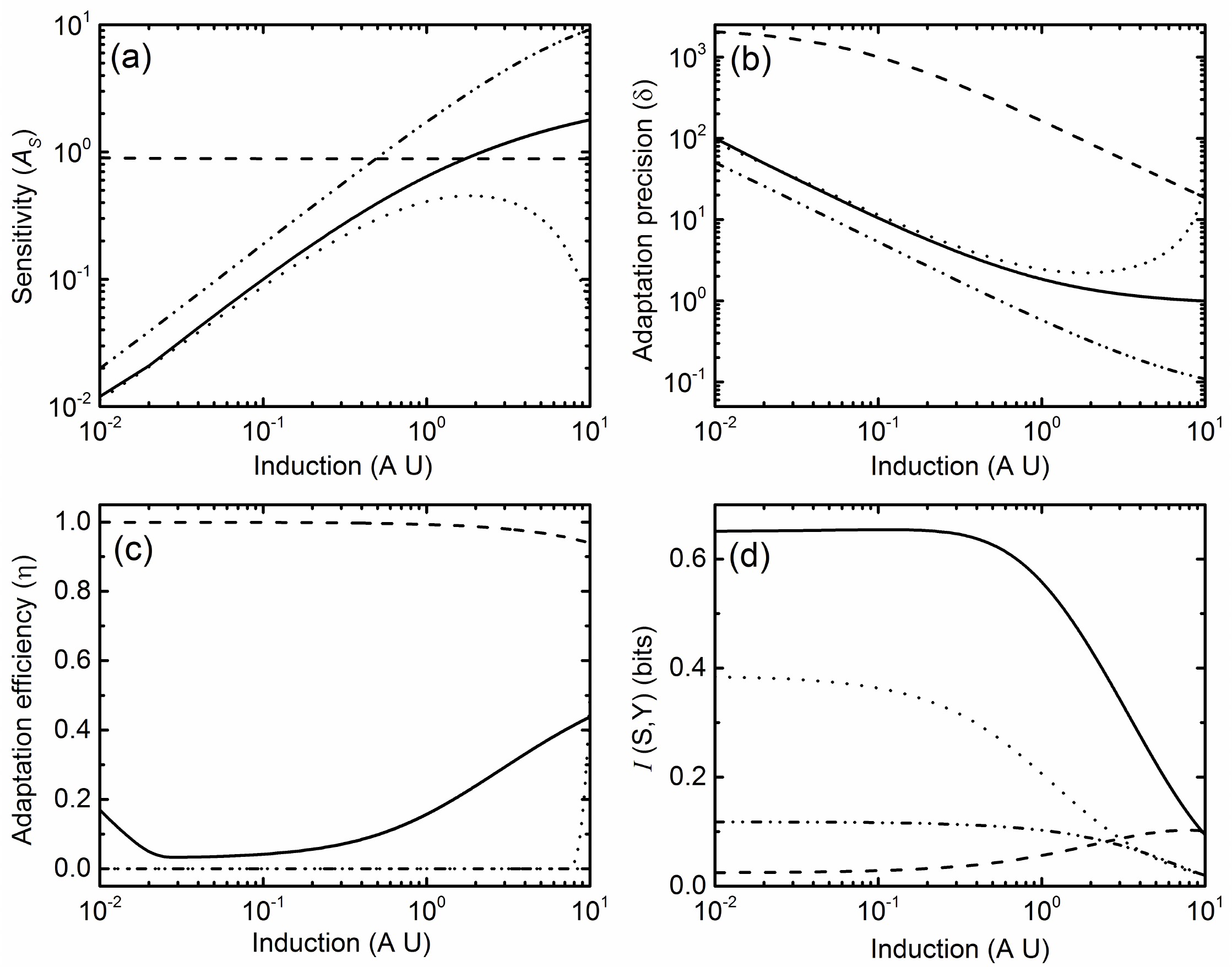}
\caption{\textbf{Performances of the incoherent feed-forward-loop motifs.}
Different measures of the network properties at a range of induction level. The lines indicates for the four types of ICFFLs: type I (solid), type II (dash), type III (dot) and type IV (dash dot dot). (a) sensitivity amplification ($A_S$), (b) adaptation precision ($\delta$), (c) adaptation efficiency ($\eta$) and (d) mutual information (${\cal I}(s,y)$) are plotted as a function of constant induction level.}
\label{fig3}
\end{center}
\end{figure*}

A perfectly adaptive system is capable of restoring its pre-stimulus state very soon after the sensitive overshoot \cite{Ma2009, Jia2016}. The mechanism of adaptation is also known as desensitisation, which means the prevention of the stimulating species even if the signalling system is functionally active \cite{Behar2007}. The speciality of desensitisation is that the system need not fully shut down itself by deactivation or degradation of the system components. Fig.~\ref{fig3}(b) represents the adaptation precisions ($\delta$) for the four ICFFL motifs. It is true that the precision quality in any adaptive system falls with increasing stimulus strength \cite{Behar2007}. In the present study, all the ICFFLs also show similar direction. The type II motif, with its three negative controls, stands unique from the others. On the other hand, an opposite trend of $A_S$ is observed in $\delta$ for type I, III and IV. Type IV is the most imprecise in adaptation where it shows better sensitivity among the three. Again it establishes the prediction that the position of the negative regulatory edge ($S\rightarrow X$) in the indirect branch has less impact on Y. Here, it is important to concentrate on the high induction range as the precision of adaptation becomes weaker at high induction than the low. To examine the role of regulatory edges on precise adaptation, one has to keep the type II at the top for its good adaptive aspect all over the range of the signal. The comparative performance of type II with the others makes one to remark that a network with more negative controls may improve $\delta$.

To discuss Fig.~\ref{fig3}(c), I need to remember the derived metric, adaptation efficiency ($\eta$). A near 1.0 efficiency of a network motif characterises the motif simultaneously as sound sensitive as well as good adaptive. The $A_S$ and $\delta$ scores of type II make it efficiently adaptive circuit, providing a good $\eta$-value ($\approx 1.0$) all over the signal range. The type I yields near $50\%$ efficiency at a quite high signal, but it is weakly efficient at low signal. In contrast with type I and II, a poor adaptation efficiency is observed in type III and IV. Though type III has a sharp rise at very high signal strength, it is wise to consider them (type III and IV) as inefficient circuits in comparison with type I and II. As the metric $\eta$ is derived from $A_S$ and $\delta$, a quantitative connection is to be drawn from the characteristics of $A_S$ and $\delta$ (see Fig.~\ref{fig3}(a) and 4.3(b)). Here, the direct ($S\rightarrow Y$) regulation plays a crucial role to discriminate type I from type III because both of them show quite similar trends in Fig.~\ref{fig3}(a) and Fig.~\ref{fig3}(b). But a clear separation makes them different in the patterns of $\eta$ in Fig.~\ref{fig3}(c).

\subsection{Stochastic behavior and mutual information}

The efficient conversion of the essence of stimulus to the output of the circuits possess significance due to the transmission of information. The biological information processing is always probabilistic due to the rare molecular population. To quantify the mutual information (${\cal I}(s,y)$) between the input node (S) and the output node (Y) in the presence of fluctuations, I have adopted both analytical and numerical approaches. As mutual information is a quantitative and more general measure than the correlation coefficients to quantitate the association between two fluctuations spaces \cite{Mehta2008}, I calculate the two-time correlation between the noise processes $\xi_s$ and $\xi_y$ to compute ${\cal I}(s,y)$. In Fig.~\ref{fig3}(d), I analyse the informational aspects for the four motifs. I know that for an effective Gaussian-type noise, the maximum information transmission occurs because of the maximum entropy property \cite{Borst1999}. Here, the approximated Gaussian channels exhibit interesting features and the most significant trends for the motivation of this paper. The type I carries maximum information up to a large dose of stimulus and decays at very high level. Though it is shown in reported articles that the ICFFLs cannot reduce signalling noise \cite{Ghosh2005}, the type I and III ICFFL must possess some features to carry better information than the others. Oppositely both type II and IV are weak in information transmission throughout the dose range. As an explanation, it can be said that a signalling circuit, processing satisfactorily better information, must have the efficiency to discriminate between different input dose levels. To correlate the information flow and network structure, the presence of positive and negative regulatory edges are significant here. On a close observation of the four ICFFLs, one can remark that type II transmits weak information for the presence of three inhibitory regulations. It is also important to note that the $S\rightarrow X$ inhibition reduces the quality of information. Though information loss occurs on the elongation of cascade lengths \cite{Mehta2008,Tkacik2011}, it will not be evident for the current study. The observation presents all the three node circuits with different regulations transferring different pieces of information.

\subsection{Evolutionary trade-offs}

\begin{figure*}[h]
\begin{center}
\includegraphics[width=0.7\linewidth,angle=0]{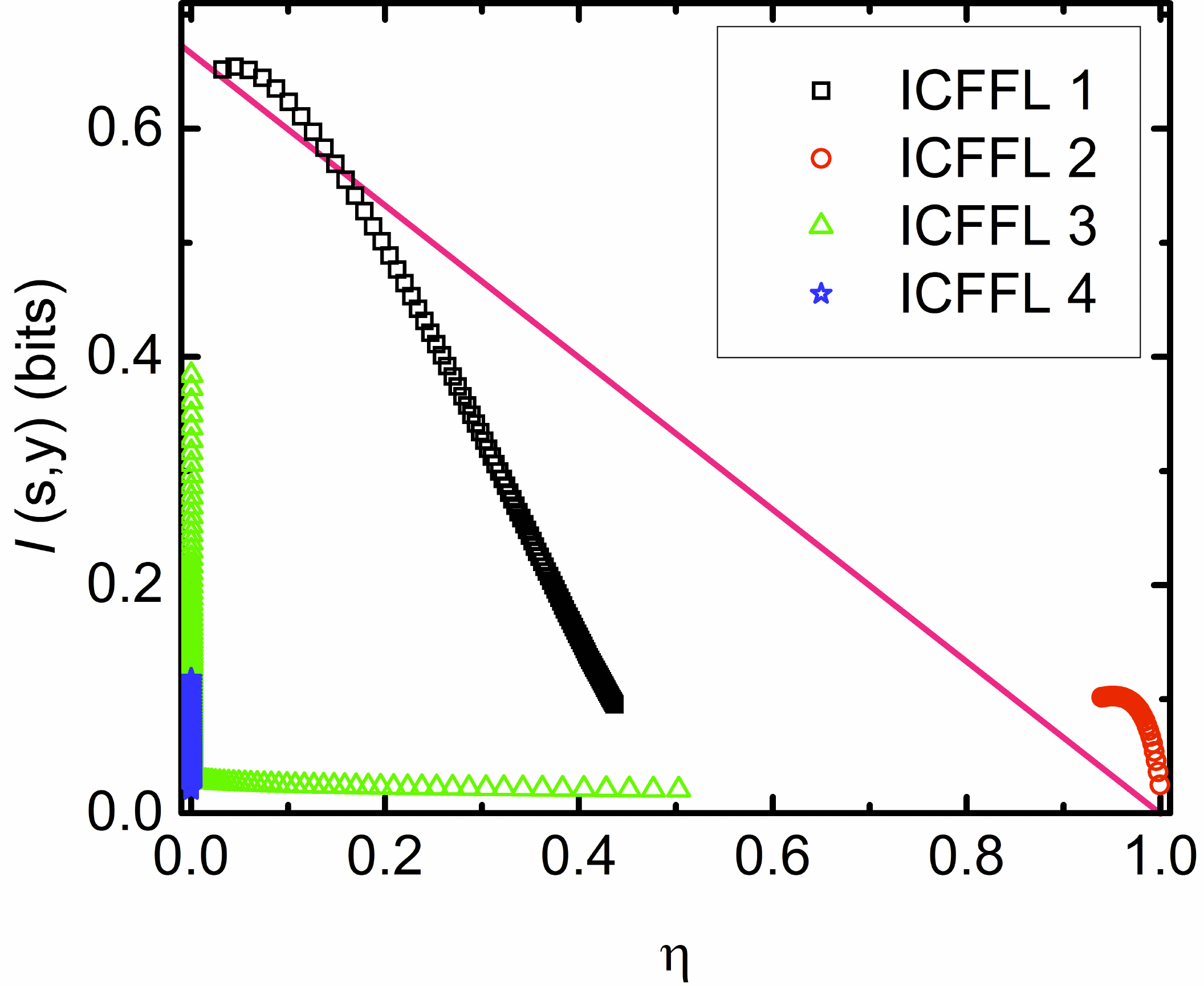}
\caption{\textbf{Allocation of the ICFFLs in task space and the Pareto front.}
All the four types of incoherent feed-forward loops are projected in the Pareto task space. The pink line is the Pareto front for the optimal task performances along the two axes ($\eta$ and ${\cal I}(s,y)$). All the points are generated for the four ICFFLs: type I (black square), type II (red circle), type III (green triangle) and type IV (blue star) at the range of the induction.}
\label{fig4}
\end{center}
\end{figure*}

The theoretical experiment of the current paper is designed to reach a decisive state about the optimality of the four ICFFLs. The temporal profiles of the four types are analysed to calculate sensitivity amplification and adaptation precision. These enable one to find out the adaptation efficiency for a particular dose of stimulus, and from the stochastic study, the mutual information has been quantified. Here, the four ICFFLs motifs are found to be multitasking. They show sensitive, adaptive as well as informative features, which are necessary for a biochemical motif to be potent in the way of natural selection. But any of the four motifs can not perform well in all the tasks. So there must be some trade-offs among the tasks. To address the trade-offs, I prefer to employ the concept of Pareto efficiency. I construct a two-dimensional Pareto task space considering the mutual information (${\cal I}(s,y)$) along the ordinate and adaptation efficiency ($\eta$) along the abscissa as two different tasks (see Fig.~\ref{fig4}). The maxima of the performances of the two tasks by all the motifs are connected to get the Pareto front (pink line). Any point near the front in the task space is said to be more Pareto efficient than the distal points. Also, a point near the terminals of the front is more specialist in that particular task than a generalist in two tasks. A generalist possesses more favour than a specialist in evolution. I depict the set of ${\cal I}(s,y)$ and $\eta$ performances for all the motifs at a range of the stimulus induction. To observe the optimality of a particular motif in the task space, I have to remark that ICFFL I has higher fitness than the others because, its performance in both the tasks, ${\cal I}(s,y)$ and $\eta$, is satisfactory as a generalist. Comparatively, ICFFL II, as a good specialist in adaptation does not hold a good position in the evolutionary selection. To discuss the performances of type III and IV, I have to say that they are the poor performers in the Pareto task space to become less fit in evolution. The reported abundance of the four types in nature \cite{Mangan2003a} supports the results of the current observation.

\section{Conclusion}

The present paper encounters the variety of signal utilisation in the four types of incoherent feed-forward loop motifs. The kinetic model for each type has been simulated for the same range of stimulus, which induces the synthesis of the motif input, $S$. Keeping all other rate parameters identical, I quantify four different aspects of signal transduction. From the quantified results, I make an observed prediction that in any three-node motif, if the number of negative regulatory edges increase, the motif shows more precise adaptation. On the other hand, the motif also will be sparse in information processing. In the current study, type II is of that particular type among the four motifs.

In reported literature on noise characterization in FFLs, the noise reduction in the coherent FFLs and also their inconsistent abundance have been explained \cite{Mangan2003a, Ghosh2005}. But for incoherent FFLs, the type I motif, which is not least noisy but most abundant, has not been explained clearly. In the present study, I have successfully demonstrated the trend in the ICFFLs abundance using Pareto optimality. In Pareto task space, I find type I incoherent motif as most efficient among the four ICFFLs. Here, I may comment that a motif, to perform efficiently in biochemical cascades, need not be least noisy. Again, as the fluctuations are manifested around the mean field dynamics, the deterministic characterization of the adaptation efficiency does not conflict with the stochastic characterization of the mutual information and the projection of the adaptation efficiency and mutual information along task performance axes in multi-objective task space is justified.

\section{Acknowledgements}
I am thankful to Alok Kumar Maity and Arnab Bandyopadhyay for critical reading of the manuscript. TM acknowledges financial support from CSIR, India [01(2771)/14/EMR-II].


\clearpage

\begin{table*}[h!]
\centering
\caption{
List of kinetics schemes and rate parameters used in the model.
} 
{\begin{tabular}{lllll||}
\hline
{Description} & {Reaction} & {Propensity function}& {Rate constant} \\
\hline
Basal synthesis of S			&	$\phi \longrightarrow S$				&	$k_0$			&	$k_0=0.005\ \text{ molecules\ min}^{-1}$\\
Induced synthesis of S		&	$I \longrightarrow S$ 				&	$k_1I$			&	$k_1=0.01\ \text{min}^{-1}$\\
Degradation of S			&	$S \longrightarrow \phi$				&	$\tau^{-1}_ss$		&	$\tau_s=10\ \text{min}$\\
S mediated synthesis of X		&	$\phi \xrightarrow{f_{x}(s)} X$		&	$k_2f_x(s)$		&	$k_2=1.7\ \text{ molecules\ min}^{-1}$\\
Degradation of X			&	$X \longrightarrow \phi$				&	$\tau^{-1}_xx$		&	$\tau_x=10\ \text{min}$\\
S \& X mediated synthesis of Y	&	$\phi \xrightarrow{f_{y}(s)g_{y}(x)}$	&	$k_3f_{y}(s)g_{y}(x)$&	$k_3=29.0\ \text{ molecules\ min}^{-1}$\\
Degradation of Y			&	$Y \longrightarrow \phi$				&	$\tau^{-1}_yy$		&	$\tau_y=10\ \text{min}$\\
\hline
\end{tabular}}
\label{table1}
\end{table*}

\begin{table*}[h!]
\centering
\caption{
List of the functions used in Table I.
} 
{\begin{tabular}{lllllccccc}
\hline
Function &		 Type I	& 	Type II 			&	 Type III		&	 Type IV \\
\hline
$f_x(s)$	&	\Large$\frac{s}{s+K_{xs}}$	&	\Large$\frac{K_{xs}}{s+K_{xs}}$	&	\Large$\frac{s}{s+K_{xs}}$	&	\Large$\frac{K_{xs}}{s+K_{xs}}$\\
$f_y(s)$	&	\Large$\frac{s}{s+K_{ys}}$	&	\Large$\frac{K_{ys}}{s+K_{ys}}$	&	\Large$\frac{K_{ys}}{s+K_{ys}}$	&	\Large$\frac{s}{s+K_{ys}}$\\
$g_y(x)$	&	\Large$\frac{K_{yx}}{x+K_{yx}}$	&	\Large$\frac{K_{yx}}{x+K_{yx}}$	&	\Large$\frac{x}{x+K_{yx}}$	&	\Large$\frac{x}{x+K_{yx}}$\\

\hline
\end{tabular}}
\label{table2}
\end{table*}


\begin{thebibliography}{99}
\providecommand{\url}[1]{\texttt{#1}}
\providecommand{\urlprefix}{URL }
\expandafter\ifx\csname urlstyle\endcsname\relax
  \providecommand{\doi}[1]{doi:\discretionary{}{}{}#1}\else
  \providecommand{\doi}{doi:\discretionary{}{}{}\begingroup
  \urlstyle{rm}\Url}\fi
\providecommand{\bibAnnoteFile}[1]{%
  \IfFileExists{#1}{\begin{quotation}\noindent\textsc{Key:} #1\\
  \textsc{Annotation:}\ \input{#1}\end{quotation}}{}}
\providecommand{\bibAnnote}[2]{%
  \begin{quotation}\noindent\textsc{Key:} #1\\
  \textsc{Annotation:}\ #2\end{quotation}}
\providecommand{\eprint}[2][]{\url{#2}}

\bibitem{Shoval2010} Shoval, O., Goentoro, L., Hart, Y., Mayo, A., Sontag, E. and Alon, U., 2010. Proceedings of the National Academy of Sciences, 107(36), pp.15995-16000.

\bibitem{Macnab1972} Macnab, R.M. and Koshland, D.E., 1972. Proceedings of the National Academy of Sciences, 69(9), pp.2509-2512.

\bibitem{Koshland1982} Koshland Jr, D.E., Goldbeter, A. and Stock, J.B., 1982. Science, 2, p.16.

\bibitem{Yi2000} Yi, T.M., Huang, Y., Simon, M.I. and Doyle, J., 2000. Proceedings of the National Academy of Sciences, 97(9), pp.4649-4653.

\bibitem{Hansen2008} Hansen, C.H., Endres, R.G. and Wingreen, N.S., 2008. PLoS Comput Biol, 4(1), p.e1.

\bibitem{Endres2006} Endres, R.G. and Wingreen, N.S., 2006. Proceedings of the National Academy of Sciences, 103(35), pp.13040-13044.

\bibitem{Miller1986} Miller, K.J., Kennedy, E.P. and Reinhold, V.N., 1986. Science, 231(4733), pp.48-51.

\bibitem{Reisert2001} Reisert, J. and Matthews, H.R., 2001. The Journal of physiology, 530(1), pp.113-122.

\bibitem{Stadel1983} Stadel, J.M., Nambi, P., Shorr, R.G., Sawyer, D.F., Caron, M.G. and Lefkowitz, R.J., 1983. Proceedings of the National Academy of Sciences, 80(11), pp.3173-3177.

\bibitem{Tsimring2014} Tsimring, L.S., 2014. Reports on Progress in Physics, 77(2), p.026601.

\bibitem{Tostevin2010} Tostevin, F. and Ten Wolde, P.R., 2010. Mutual information in time-varying biochemical systems. Physical Review E, 81(6), p.061917.

\bibitem{Alon2006} U. Alon, An Introduction to Systems Biology: Design Principles of Biological Circuits (CRC Press, 2006).

\bibitem{Mangan2003a} Mangan, S. and Alon, U., 2003. Proceedings of the National Academy of Sciences, 100(21), pp.11980-11985.

\bibitem{Mangan2003b} Mangan, S., Zaslaver, A. and Alon, U., 2003. Journal of molecular biology, 334(2), pp.197-204.

\bibitem{Mangan2006} Mangan, S., Itzkovitz, S., Zaslaver, A. and Alon, U., 2006. Journal of molecular biology, 356(5), pp.1073-1081.

\bibitem{Kalir2005} Kalir, S., Mangan, S. and Alon, U., 2005. Molecular systems biology, 1(1).

\bibitem{Kaplan2008} Kaplan, S., Bren, A., Dekel, E. and Alon, U., 2008. Molecular systems biology, 4(1), p.203.

\bibitem{Dekel2005} Dekel, E. and Alon, U., 2005. Nature, 436(7050), pp.588-592.

\bibitem{Milo2002} Milo, R., Shen-Orr, S., Itzkovitz, S., Kashtan, N., Chklovskii, D. and Alon, U., 2002. Science, 298(5594), pp.824-827.

\bibitem{Kremling2008} Kremling, A., Bettenbrock, K. and Gilles, E.D., 2008. Bioinformatics, 24(5), pp.704-710.

\bibitem{Wall2005} Wall, M.E., Dunlop, M.J. and Hlavacek, W.S., 2005. Journal of molecular biology, 349(3), pp.501-514.

\bibitem{Guo2009} Guo, D. and Li, C., 2009. Physical Review E, 79(5), p.051921.

\bibitem{Ghosh2005} Ghosh, B., Karmakar, R. and Bose, I., 2005. Physical biology, 2(1), p.36.

\bibitem{Bose2004} Bose, I., Ghosh, B. and Karmakar, R., 2005. Physica A: Statistical Mechanics and its Applications, 346(1), pp.49-57.

\bibitem{Goentoro2009} Goentoro, L., Shoval, O., Kirschner, M.W. and Alon, U., 2009. Molecular cell, 36(5), pp.894-899.

\bibitem{Briat2016} Briat, C., Gupta, A. and Khammash, M., 2016. Cell systems, 2(1), pp.15-26.

\bibitem{Goldbeter1984} Goldbeter, A. and Koshland, D.E., 1984. Journal of Biological Chemistry, 259(23), pp.14441-14447.

\bibitem{Huang1996} Huang, C.Y. and Ferrell, J.E., 1996. Proceedings of the National Academy of Sciences, 93(19), pp.10078-10083.

\bibitem{Ma2009} Ma, W., Trusina, A., El-Samad, H., Lim, W.A. and Tang, C., 2009. Cell, 138(4), pp.760-773.

\bibitem{Jia2016} Jia, C. and Qian, M., 2016. PloS one, 11(5), p.e0155838.

\bibitem{Behar2007} Behar, M., Hao, N., Dohlman, H.G. and Elston, T.C., 2007. Biophysical journal, 93(3), pp.806-821.

\bibitem{Mehta2008} Mehta, P., Goyal, S. and Wingreen, N.S., 2008. Molecular systems biology, 4(1), p.221.
  
\bibitem{Shannon1948} Shannon, C. E., 1948. Bell Syst Tech J, 27, pp.623.

\bibitem{Kampen2007} N. G. van Kampen, Stochastic Processes in Physics and Chemistry, 3rd ed (North-Holland, 2007).

\bibitem{Paulsson2005} Paulsson, J., 2005. Physics of life reviews, 2(2), pp.157-175.

\bibitem{Maity2014} Maity, A.K., Bandyopadhyay, A., Chaudhury, P. and Banik, S.K., 2014. Physical Review E, 89(3), p.032713.

\bibitem{Elf2003} Elf, J. and Ehrenberg, M., 2003. Genome research, 13(11), pp.2475-2484.

\bibitem{Swain2004} Swain, P.S., 2004. Journal of molecular biology, 344(4), pp.965-976.

\bibitem{Paulsson2004} Paulsson, J., 2004. Nature, 427(6973), pp.415-418.

\bibitem{Gardiner2009} C. W. Gardiner, Stochastic Methods, 4th ed (Springer, 2009).

\bibitem{Cover2012} Cover, T.M. and Thomas, J.A., 2012. Elements of information theory. John Wiley \& Sons.

\bibitem{Gillespie1976} Gillespie, D.T., 1976. Journal of computational physics, 22(4), pp.403-434.

\bibitem{Gillespie1977} Gillespie, D.T., 1977. The journal of physical chemistry, 81(25), pp.2340-2361.

\bibitem{Noor2012} Noor, E. and Milo, R., 2012. Science, 336(6085), pp.1114-1115.

\bibitem{Shoval2012} Shoval, O., Sheftel, H., Shinar, G., Hart, Y., Ramote, O., Mayo, A., Dekel, E., Kavanagh, K. and Alon, U., 2012. Science, 336(6085), pp.1157-1160.

\bibitem{Sheftel2013} Sheftel, H., Shoval, O., Mayo, A. and Alon, U., 2013. Ecology and evolution, 3(6), pp.1471-1483.

\bibitem{Pareto1971} Pareto, V., 1971. Manual of political economy. Augustus M. Kelley, New York.

\bibitem{Borst1999} Borst, A. and Theunissen, F.E., 1999. Nature neuroscience, 2(11), pp.947-957.

\bibitem{Tkacik2011} Tka{\v c}ik, G. and Walczak, A.M., 2011. Journal of Physics: Condensed Matter, 23(15), p.153102.

\end{thebibliography}
\end{document}